# Two regimes in conductivity and the Hall coefficient of underdoped cuprates in strong magnetic fields

L P Gor'kov[1,2]  and  G B Teitel'baum[3]

[1] NHMFL, Florida State University, 1800 East Paul Dirac Drive, Tallahassee Florida 32310,    USA
[2] L.D. Landau Institute for Theoretical Physics of the RAS, Chernogolovka 142432, Russia
[3] E.K. Zavoiskii Institute for Technical Physics of the RAS, Kazan 420029, Russia

We address recent experiments shedding light on the energy spectrum of under- and optimally doped cuprates at temperatures above superconducting transition. Angle resolved photoemission reveals coherent excitation only near nodal points on parts of the "bare" Fermi surface known as the Fermi arcs. The question debated in the literature is whether the small normal pocket, seen via quantum oscillations exists at higher temperatures or forms below a charge order transition in strong magnetic fields. Assuming the former case as possibility,  expressions derived for the resistivity and the Hall coefficient (in weak and strong magnetic fields) with both types of carriers participating in transport. There are two regimes. At higher temperatures (at a fixed field) electrons are dragged by the Fermi arcs' holes. The pocket being small its contribution into conductivity and the Hall coefficient is negligible. At lower temperatures electrons decouple from holes behaving as a Fermi gas in the magnetic field. Mobility of holes on the arcs decreasing in strong fields with decrease of temperature,  below a crossover point the pocket electrons prevail  changing sign of the Hall coefficient in the low temperature limit. Such behavior finds its confirmation in recent high-field experiments.

## 1. Introduction

Understanding high temperature superconductivity (HTSC) in cuprates is the pressing issue both on the theory side of the problem and for practical implications. While in superconductors of "old" generation it was at least known that superconductivity emerges from the normal Fermi liquid phase, HTSC is preceded by the so-called pseudogap phase with many abnormal properties. Most unexpected feature revealed in the angle resolved photoemission (ARPES) experiments is that the coherent electronic excitations exist only on the Fermi arcs along the "bare" Fermi surface separated from each other by large energy gaps [1].

It was absolutely unclear how to interpret even such basic experimental data as for resistivity or the Hall coefficient. For the first time the importance of the question was probably realized when it turned out that the Hall coefficient [2] in $La_{1-x}Sr_xCuO_4$ (LSCO) manifests the activation temperature dependence [3].

In 2007 discovery of the quantum oscillations (QOs) in underdoped (UD) *ortho II* YBCO [4] has shown presence of small Fermi liquid (FL) pocket(s). Low frequency QOs in high magnetic fields were found in Hg1201 as well [6].

So far, ARPES experiments were performed on he two-layer $Bi_2Sr_2CaCu_2O_8$ (Bi2212), single-layer Bi2201 and LSCO (see recent review [7]). As yet, ARPES is not available for Hg1201 and YBCO because of problems with the surface [8]. The consensus is that, except for minor details, ARPES findings bear the general character and reflect the basic physics.

It is all the more remarkable that the recent experiments [9] added new and convincing evidence in support of the Fermi arcs concept. In brief, it was pointed out in [9] that characteristic for temperature dependence of resistivity of the clean Hg1201 (and few others cuprate compounds) is the contribution proportional to square of the temperature, as if in a Fermi liquid. In [9] the latter was related to free carriers on the Fermi arcs.

In turn, non-monotonous temperature and field dependence of the Hall and Seebeck coefficients in YBCO and Hg1201 in experiments [11, 12] below $T \approx 100-150 K$ was considered as due to contribution from electrons on the pocket.

These results pose question about interrelation between pocket(s) and the Fermi arcs. Currently, one of highly debated issue in the literature concerns nature of the pocket, that is, whether the pocket is a band feature and is omnipresent in the spectrum, including higher temperatures, or it emerges below a charge order (CDW) transition. As to the latter, according to [13], a CDW ordering indeed occurs in the relevant intervals of temperature and the magnetic field.

Let us emphasize that most QO experiments were performed in high fields and not too high temperatures ($T < 10$ K ). We investigate conductivity and the Hall coefficient assuming that carriers of both type participate in the transport properties in all temperature range of the pseudogap phase. We adopt the view [5] that only one small pocket exists in UD YBCO that lies away from a Fermi arc.

The model of holes on the Fermi arcs interacting via short-range interactions was examined in frameworks of the kinetic equation in [10] for interacting holes on the Fermi arcs. Below the model is generalized to include electrons on the pocket and extended to the case of strong magnetic fields. Looking ahead, our results do not contradict [11, 12]; more importantly, in our opinion, the result give support to the notion of such pocket as a band feature.

In its most general form the kinetic equation is:

$$\frac{dn}{dt} = I_{col}. \qquad (1)$$

On the left, the total derivative ($dn/dt$) accounts for variations of the Fermi distribution of the quasiparticles at their motion in the real and in the momentum space in the presence of external fields. The right hand side term, the collision integral $I_{col}$ is responsible for the relaxation processes.

**2. Results from the model of Fermi arcs**

Consider first carriers on the Fermi arcs. In weak constant electric and magnetic fields the kinetic equation (1) is presented in the commonly-accepted form:

$$e(\vec{E} + \frac{1}{c}[\vec{v} \times \vec{H}])\vec{\nabla}_{\vec{p}} n(\vec{p}) = I_{coll}. \qquad (2)$$

Changes $n_1$ of the Fermi distribution function $n(\vec{p})$ are also small. At a non-zero current the system of charged carriers moves as a whole with a drift velocity $\vec{u}$. Correspondingly, the left hand side (L.H.S.) in (2) reduces to:

$$e\{(\vec{E}\cdot(\vec{v}-\vec{u})) - \frac{1}{c}([\vec{v}\times\vec{H}]\cdot\vec{u})\}\frac{\partial n_o}{\partial \varepsilon}. \qquad (3)$$

(Here ($\partial n_o / \partial \varepsilon$) is the derivative of the equilibrium Fermi function: $\vec{v} = d\varepsilon(\vec{p})/d\vec{p}$). Similarly the collision integral in the right hand side (R.H.S.) of Eq.(3) should be expanded over small $n_1$. The balance between L.H.S. and R.H.S. gives expressions for the quadratic resistivity term and for the Hall coefficient, correspondingly [10]:

$$\rho_{2D}(T) = \left(\frac{\pi\hbar}{e^2}\right)\frac{4\pi^2 |Z\tilde{V}(1;2)|^2}{3(\Delta\varphi)^2}\sqrt{\frac{\Delta}{\varepsilon_F^*}}\left(\frac{T}{\varepsilon_F^*}\right)^2 \qquad (4)$$

and

$$R_H = \frac{1}{ecn_{eff}} > 0. \qquad (5)$$

Here $Z\tilde{V}(1;2)$ is the renormalized matrix element of the electron-electron interaction, $m^* = Z^{-1}m$ is the renormalized mass. The combination $\varepsilon_F^* = p_F^2/2m^*$ is defined as the renormalized Fermi energy.

In Eq. (4) $\Delta = v_F p_F[(K/4p_F) - 1]$, $K^{\parallel}$ is projection of the Umklapp vector $\vec{K} = (2\pi/a, 2\pi/a)$ on the diagonal in the Brillouin zone (BZ). Processes with the umklapp scattering are critical for relaxation of the momentum and, hence, for non-zero resistivity. Unlike ordinary metals with large Fermi surfaces, for system of the narrow Fermi arcs the resistivity was obtained in the explicit form of Eq.(4); $\Delta\varphi$ is the arc width. For the isotropic energy spectrum the effective number of carriers is: $n_{eff} = \frac{\Delta\varphi p_F^2}{\pi^2}(\frac{s}{c_z})$. Here $c_z$ is the lattice constant in the perpendicular-to-plane direction, $s$ – the number of the CuO$_2$-planes per unit cell. Note in passing that each parameter in (4, 5) can be found experimentally.

In fact, position of the nodal points $p_F$ is known from ARPES, the Fermi arc width $\Delta\varphi(x)$ follows directly from the experimental value of $R_H$. (The nodal points usually lie rather close to the $(\pm\pi/2, \pm\pi/2)$ - points [7]).

Rigorously speaking, expression (5) for the Hall coefficient applies only in the weak field regime $\omega_c \tau_{col} \ll 1$ ($\omega_c = (eH/m^*c)$) [10]. (Notation $\tau_{col}$ stands for the mean scattering time). Meanwhile, by the order of magnitude $\tau_{col}$ in (4) is: $\tau_{col} \propto \varepsilon_F / T^2$, i.e., $\tau_{col}$ is large in clean samples. For instance, at $T = 100$ K the magnetic field should be much smaller than 3-5 Tesla for (5) to apply. Data for $R_H$ in Hg1201, YBCO and LSCO in the whole temperature range were obtained only in higher fields (see [2], [11] and [12]).

## 3. The role of electronic pocket

The Hall coefficient Eq.(5) is temperature independent. Meanwhile, the notable feature in the Hall data [2,11,12] is its temperature dependence, $R_H(T)$ (hence, same for $\Delta\varphi(T)$).

Two mechanisms come to mind in regard to the *T*-dependence of the Hall Effect. The first one is a temperature dependence of the Fermi arcs' width that is seen in ARPES experiments [14-17]. For the model this means increase in number of carriers with the temperature increase.

The second one concerns role of the pocket. The latter was detected by high field quantum oscillations [4]. It is known that in the superconducting phase and *in zero fields* the pocket carriers are independent from holes paired on the arcs and contribute into the specific heat the residual *normal metal* linear term [5].

Experimentally, the temperature dependence of the Hall coefficient has a maximum. In the *high fields* experiments [11, 12] $R_H(T)$ even changes the sign at lower temperatures. This was ascribed to *electrons on a pocket* for which mobility is increasing as $T \to 0$. Correspondingly, we consider resistivity and the Hall coefficient for a two-component system of the Fermi arcs' holes and the pocket's electrons, first from the side of *higher* temperatures where the classic kinetic equation approach is applicable.

The general form of a current in the dimension $d$ ($d = 2, 3$) is:

$$\vec{j}_d = 2e \int \vec{v} \frac{d^d \vec{p}}{(2\pi)^d} (\vec{p} \cdot \vec{v}_{drift}) (\frac{1}{4T}) ch^{-2}(\frac{\varepsilon_p}{2T}) \quad (6)$$

with $\vec{v}_{drift}$ -a drift velocity. Introduce notations, $\vec{v}_p$ and $\vec{u}_{FA}$ for the drift velocities of electrons and holes, respectively. Since the size of the pocket size is small [4, 5] scattering events do not involve electrons directly in the umklapp processes, so that holes on the Fermi arcs tend to "drag" electrons. An electric field pulls electrons in the opposite direction than holes. Solving the linear kinetic equation allows writing the relation between $\vec{v}_p$ and $\vec{u}_{FA}$ as follows:

$$(\vec{v}_p - \vec{u}_{FA}) \frac{1}{\tau_{eh}} + \vec{v}_p \frac{1}{\tau_{imp}} = \frac{|e|\vec{E}}{m_e} \quad . \quad (7)$$

Here $1/\tau_{eh}$ is the inverse time for relaxation of the electrons *relative to* holes ($1/\tau_{imp}$ for scattering of electrons on defects is small in YBCO and Hg1201 and is neglected below).

An estimate of the Fermi energy of electrons at low temperatures $E_{Fe}$ from data [4, 5] gives: $E_{Fe} \approx 500\,\text{K}$ (below $E_{Fe}$ and $p_{Fe}$ stand for the Fermi energy and Fermi momentum of electrons; notations $\varepsilon_F$ and $p_F$, as before, are reserved for holes). One sees that already at temperatures 100 – 150 K the electrons may be treated as the degenerate Fermi gas.

The expression for $u_{FA}$ [10] rewritten as:

$$u_{FA} = -|e|E\frac{3\Delta\varphi\varepsilon_F}{8\pi m^* |ZV(1,2)|^2 T^2}\sqrt{\frac{\varepsilon_F}{\Delta}} = -|e|E\frac{\tau_{FA}}{m^*} \quad (8)$$

defines the relaxation time of holes $\tau_{FA}(T)$:

$$\tau_{FA}(T) = \frac{3\Delta\varphi\varepsilon_F}{8\pi |ZV(1;2)|^2 T^2}\sqrt{\frac{\varepsilon_F}{\Delta}}. \quad (9)$$

Interactions between electrons and holes are due to same short-range Coulomb interactions as the interactions between holes on different Fermi arcs. Consequently, the matrix elements that determine $1/\tau_{eh}$ have same order of magnitude. This makes possible from the analysis of the kinetic equation to derive the following estimate for $1/\tau_{eh}$:

$$\frac{1}{\tau_{eh}} \approx const \times \left(\frac{m_e}{m^*}\right)\left(\frac{p_F}{p_{Fe}}\right)\frac{T^2}{\varepsilon_F}, \quad (10)$$

with a $const \sim 1$. The factor $p_F/p_{Fe} \gg 1$ in (10) shows that $1/\tau_{eh} \gg 1/\tau_{FA}$. From (7) for conductivity follows:

$$\sigma_{3D}(T) = e^2\left\{\frac{n_{eff}}{m^*} - \frac{n_e}{m_e}\right\}\tau_{FA}. \quad (11)$$

where the second term is the contribution from *electrons dragged* by holes (hence, the negative sign). In (11) $n_e$ is the electronic density. Comparison between $n_{eff} = \Delta\varphi s p_F^2/c\pi^2$, the effective density of holes on the Fermi arcs' and $n_e = (p_{Fe}^3/2\pi^2)$ shows that $n_{eff} \gg n_e$ ($m^* \sim m_e$; $p_F/p_{Fe} \gg 1$), i.e. the contribution of electrons into conductivity in this regime is very small.

In the vector notations the Hall coefficient $R_H(T)$ is defined by the relation:

$$\vec{E} = \rho\vec{J} + R_H[\vec{H}\times\vec{J}]. \quad (12)$$

Here $\rho$ is the resistivity and $\vec{J}$ - the longitudinal current. The two drift velocities $v_{p,H}$ and $u_{FA,H}$ for electrons and holes in the presence of the magnetic field were found from the kinetic equations after substituting into Eq. (3) for the Lorentz forces

$$-|e|E \Rightarrow \frac{|e|}{c} H u_{FA} \; ; \text{(for holes)} \qquad (13a)$$

$$|e|E \Rightarrow -\frac{|e|}{c} H v_p \; ; \text{(for electrons)} \qquad (13b)$$

Calculations of the Hall currents are similar to that ones for the two electrical currents. The Hall coefficient of the two-component system is:

$$R_H(T) = \frac{1}{c|e|n_{eff}} \times \left\{ 1 + \left(\frac{n_e}{n_{eff}}\right)\left(\frac{m^*}{m_e}\right) \right\} . \qquad (14)$$

(Note that the small contributions from electrons in (11) and (14) have different signs).

**4. Interplay between holes and electrons in strong magnetic fields**

As mentioned above, formally, there are the restrictions on strengths of the magnetic field. For holes it reads $\omega_c \tau_{FA} \propto (\omega_c/T)(\varepsilon_F/T) \ll 1$.

The arguments are given below that expression (5) applies for the Hall coefficient of holes even if that restriction were not fulfilled. At the same time, for electrons the restriction $\omega_{ce}\tau_{eh} \ll 1$ ($\omega_{ce} \equiv (eH/m_e c)$) at the *fixed* value of the magnetic field is violated at temperatures considerably *lower* than for carriers on the Fermi arcs due to the factor $p_{Fe}/p_F \ll 1$ in $\omega_{ce}\tau_{eh} \approx \omega_{ce}(p_{Fe}/p_{Fh})\tau_{FA}$. Thus, at high enough temperatures the contribution from holes on the Fermi arcs prevails over that one from electrons.

Thereby, while holes on the Fermi arc can already be in the strong field regime for *electrons* the new physics sets in only at $\omega_{ce}\tau \approx \omega_{ce}(p_{Fe}/p_{Fh})\tau_{FA} \approx 1$. Below the pocket *decouples* from holes on the Fermi arcs and its contribution to $R_H(T)$ finally become the one of the *free* electrons in a magnetic field and scattering on impurities finally becomes the only mechanism of relaxation. To the best of the authors' knowledge no rigorous theoretical expression for $R_H(T)$ is available in general except in the limit of *strong* magnetic field.

For the diagonal (symmetric) component of conductivity in high fields estimates give [18] $\sigma^S \sim \sigma_0(\omega_c\tau)^{-2}$, where $\sigma_0$ is conductivity in the absence of the field. For holes on the Fermi arcs $\sigma^s_{FA} \sim (n_{eff}e^2/\omega_h^2 m^*)(T^2/\varepsilon_F)$, while for electrons $\sigma^s_p \sim (n_e e^2/\omega_{cp}^2 m_p)\tau_{imp}^{-1}$. One sees directly that at low temperatures the electronic mobility will exceed mobility of carriers on the Fermi arcs. Comparison of the two expressions defines the temperature scale $T_0$: $T_0^2 \approx (n_e/n_{eff})(\varepsilon_F/\tau_{imp})$ (by the order of magnitude it can be rewritten as $T_0^2 \approx (p_{Fe}/p_{Fh})(E_{Fe}/\tau_{imp}))$.

At lower temperatures the Hall coefficient is:

$$R_H(T) = -\frac{1}{c|e|n_e} \quad (15)$$

Rewriting $\omega_c \tau \approx \omega_c (p_{Fe}/p_{Fh})\tau_{FA} \approx 1$ as $T^2 \approx (p_{Fe}/p_{Fh})\omega_c \varepsilon_F$ defines the approximate position of the Hall coefficient maximum:

$$T_{max} \approx \sqrt{(p_{Fe}/p_F)\omega_{ce}\varepsilon_F} . \quad (16)$$

### 5. Hall coefficient for holes on Fermi arcs

Return to holes on the Fermi arcs in the opposite limit $\omega_c \tau_{col} \gg 1$ and show that expression (5) for $R_H(T)$ is correct in limits of both weak and strong magnetic fields.

Assuming the tetragonal symmetry for the CuO$_2$-plane in cuprates, the Hall coefficient in strong fields can be defined as $R_H H \equiv 1/\sigma_{xy}$ (below we use the system of coordinates with the $x, y$-axes along the two diagonals of the BZ). For normal metals in the strong field limit the value of the non-diagonal conductivity component $\sigma_{xy}$ is known to depend critically on whether electrons at the Fermi surfaces in the magnetic field move along closed or open trajectories (see e. g. [18]). In particular, in the former case:

$$\sigma_{xy} = -\frac{|e|c}{H}n. \quad (17)$$

(Here $n$ is the number of carriers for a closed Fermi surface).

Returning to Eq. (1), in strong magnetic fields $\omega_c \tau_{col} \gg 1$ the collision integrals obviously should be omitted. Finding the distribution functions for carriers placed in the strong magnetic field and a weak electric field reduces to solving the equations of motion, as given by the Liouville's Theorem:

$$\frac{dn}{dt} = 0 . \quad (18)$$

The approach to solving (18) was elaborated in [19]. (See summary of the method in [18]).

Unlike weak fields, as in Eq.(3), the charges now move along trajectories bent by the strong magnetic field. Because of that presenting the distribution functions in terms of the Cartesian momentum components, $p_x, p_y$ is not convenient. Instead, in [19] it was suggested to go over to the new variables: energy $\varepsilon$ and the time variable $\tau$ defined via the relation:

$$d\tau = \frac{ds}{v_\perp}(c/eH) \quad (19)$$

(Here $ds$ is the element of "length" along the Fermi surface: $ds^2 \equiv dp_x^2 + dp_y^2$ ; $v_\perp^2 = v_x^2 + v_y^2$). Eq. (8) means that in a strong field, $H$ carriers move along the Fermi surface with high speed.

For a weak *electrical* field Eq. (18) takes the form [18,19]:

$$\frac{dn}{dt} \approx -\frac{dn_0}{d\varepsilon}e(\vec{v}\cdot\vec{E}) + \frac{\partial n_1}{\partial \tau} \approx 0. \quad (20)$$

(The collision integrals were discarded). As to $n_1$, the latter is presented [18,19] in the form:

$$n_1 = (\partial n_0 / \partial \varepsilon)e(\vec{E}\cdot\vec{g}). \quad (21)$$

The equation for $\vec{g}$ is:

$$\partial \vec{g} / \partial \tau = \vec{v}. \quad (22)$$

It is important that solution of (22) cannot contain a constant vector $\vec{g}_0$. (Such term would result in a meaningless constant shift of the distribution function (21)). The further calculations for the Fermi arcs model are, in some sense, even simpler than in [19].

In fact, there is no dilemma now concerning motion along either open or closed trajectories in the momentum space: carriers cannot go away from arcs because on both sides of the arc the distribution function is restricted by large energy gaps. In [18] the "time" $\tau$ for a carrier was formally defined by its initial position on the Fermi surface. The actual physical meaning of such procedure [19] is that it indirectly characterizes carriers by their velocities $ds/d\tau$ along the Fermi surface; as follows from Eq. (19). The Fermi arcs being separated from each other by large energy gaps where the distribution function is zero, the time variables $\tau$ of Eq. (19) can be introduced for each Fermi arc *independently*.

One can now but refers to Eq. (84, 11) in [18]. The latter reads now as the sum over the four arcs:

$$\sigma_{\alpha\beta} = \frac{2e^3 H}{(2\pi)^2 c}\sum \int_{\Delta\varphi_i} v_\alpha g_\beta d\tau dp_z \equiv \frac{2e^3 H}{\pi^2 c}\int_{\Delta\varphi} v_\alpha g_\beta d\tau dp_z, \quad (23)$$

(Sign $\Delta\varphi$ at the integral reminds that integration in (23) is limited by one Fermi arc). For definiteness, chose $\sigma_{xy}$:

$$\sigma_{xy} = \frac{2e^3 H}{\pi^2 c}\int_{\Delta\varphi} v_x g_y d\tau dp_z. \quad (24)$$

After writing out the equation

$$\frac{d\vec{p}}{d\tau} = \frac{e}{c}[\vec{H}\times\vec{v}] \quad (25)$$

in the $x, y$-components, solving Eq.(11) gives $g_y = -(c/eH)p_x$ while $v_x = (c/eH)(dp_y/d\tau)$; one finds that integration over $\tau$-variable in (24) is actually integration over the momentum component $p_y$:

$$\sigma_{xy} = \frac{2e^3 H}{\pi^2 c}\int_{\Delta\varphi} v_x g_y d\tau dp_z \equiv -\frac{2ec}{\pi^2 H}\left(\frac{s}{c_z}\right)\int_{\Delta\varphi} p_x dp_y, \quad (26)$$

where the last integral is over the area $\delta S$ limited by the single arc:

$$\sigma_{xy} = -\frac{ec}{H} n_{eff}. \qquad (27)$$

In the isotropic case $\delta S = (p_F^2/2)\Delta\varphi$ and $n_{eff} = (\Delta\varphi)p_F^2 s / c_z \pi^2$ is defined as in Eq. (5). At $\Delta\varphi \Rightarrow \pi/2$ one gradually returns to the expression of the Hall constant for the closed Fermi surface [18]. In the anisotropic case the number of carriers is expressed in terms of the area $\delta S$ limited by the single arc. (For holes: $(-e) \Rightarrow |e|$. One sees that introduction of the time variable $\tau$ for the *finite* Fermi arc was nothing more than a convenient mathematical device). Eqs. (26, 27) conclude the proof.

## 6. Conclusions

In summary, the expressions for resistivity and the Hall coefficient were derived for two types of carriers participating in transport. There are two regimes depending on strength of the magnetic field. At a fixed magnetic field and at higher temperatures electrons on the pocket are bound to holes on the Fermi arcs. Since size of the pocket is small, its contribution into conductivity and the Hall coefficient is negligible.

At lowering the temperature electrons decouple from holes and behave as a Fermi gas in the magnetic field. In the limit of low temperatures namely electrons contribute most to the Hall Effect changing sign of the Hall coefficient. The result seems to agree with recent high-field experiments.

Contributions from holes on the Fermi arcs at higher temperatures prevail and the Hall coefficient $R_H(T)$ determines the actual number of carriers on the arcs. Its decrease with increase of temperature reflects the increase of the widths of the Fermi arcs $\Delta\varphi(T)$.

At higher temperatures the contribution into $R_H(T)$ from holes on the Fermi arcs is proven to be rigorous in both limits of weak and strong magnetic fields. In the expression for the Hall coefficient the Fermi arcs width can be not small.

As to the nature of the electronic pocket, the results are consistent with the notion of its being a particular feature in the spectrum.

## 7. Acknowledgements

The work of L.P.G. was supported by the NHMFL through NSF Grant No. DMR-1157490, the State of Florida and the U.S. Department of Energy; that of G.B.T. by the Russian Academy of Sciences through Grants No. P19 and No.OFN 03.